\documentclass[aps,prb,twocolumn,showpacs,superscriptaddress,amsmath,amssymb]{revtex4-1}

\usepackage{braket}			% Allows use of Dirac notation
\usepackage{graphicx}		% Include figure files
\usepackage{color}		% Allows \textcolor{red}{Hello world}
\usepackage[normalem]{ulem}	% Allows strikethrough \sout{text goes here}
\usepackage{amsmath}

\begin{document}

%\preprint{}

\title{An Optical Spin Read-out Method for a Quantum Dot using the AC Stark Effect}


\author{Edward B. Flagg}
\email[]{edward.flagg@mail.wvu.edu}
%\thanks{}
\affiliation{Department of Physics and Astronomy, West Virginia University, Morgantown, WV 26506}

\author{Glenn S. Solomon}
%	\email{solomon@jqi.umd.edu}
	\affiliation{Joint Quantum Institute, National Institute of Standards and Technology, 
				\& University of Maryland, Gaithersburg, MD, USA.}


\date{\today}

\begin{abstract}
We propose a method to read-out the spin-state of an electron in a quantum dot in a Voigt geometry magnetic field using cycling transitions induced by the AC Stark effect.  We show that cycling transitions can be made  possible by a red-detuned, circularly-polarized laser, which modifies the spin eigenstates and polarization selection rules {\it via} the AC Stark effect.  A Floquet-Liouville supermatrix approach is used to calculate the time-evolution of the density matrix under the experimental conditions of a spin read-out operation.  With an overall detection efficiency of 2.5\%, the read-out is a single-shot measurement with a fidelity of 76.2\%. 
\end{abstract}

% 73.21.La Quantum dots, electron states and collective excitation in
% 42.50.-p Quantum optics
% 42.50.Hz Dynamic Stark shift
% Extra numbers, not used:
% 42.50.Ex Quantun information, optical implementations
% 71.70.Ej Stark effect in condensed matter

%\keywords{}

\maketitle

\section{\label{Intro}Introduction}
Quantum information science holds great promise in the areas of secure communication and rapid computation, but the physical components of a future quantum computer are still a work in progress. Any physical realization of a quantum bit, or qubit, requires several different single-qubit operations: initialization, manipulation, and read-out of its quantum state \cite{divincenzo_physical_2000}. There are many candidate systems that may act as qubits, including the spin degree of freedom of a single electron or hole trapped in a quantum dot (QD) \cite{warburton_single_2013}. Optically active transitions to many-body excited states allow the spin of the single particle ground state to be influenced by external application of oscillating electric fields, such as lasers. Spin initialization has been accomplished in a number of experimental situations involving a magnetic field in either the Faraday configuration, where the magnetic field is aligned parallel to the optical axis,
\cite{kroutvar_optically_2004,atature_quantum-dot_2006} 
or the Voigt configuration, where the magnetic field is aligned orthogonal to the optical axis 
\cite{bracker_optical_2005,xu_fast_2007,gerardot_optical_2008,carter_quantum_2013,press_complete_2008}. 
A magnetic field in the Voigt configuration allows both spin initialization and coherent manipulation because the field modifies the polarization selection rules of the optical transitions 
\cite{imamoglu_quantum_1999,berezovsky_picosecond_2008,press_complete_2008}.
Statistically significant spin read-out has been achieved in both Voigt and Faraday configurations 
\cite{berezovsky_nondestructive_2006,mikkelsen_optically_2007,atature_observation_2007,xu_fast_2007,heiss_charge_2008,press_complete_2008,lu_direct_2010,vamivakas_spin-resolved_2009},
but the lack of a cycling transition in the Voigt configuration makes a single-shot read-out of the spin-state very difficult. 

A single-shot measurement determines the state of the qubit faster than the back-action of the measurement disturbs the state. In a charged QD in a Voigt magnetic field, there is no optical transition that would leave the electron state unchanged with high fidelity 
\cite{xu_fast_2007}.
In contrast, the Faraday magnetic field configuration results in cycling transitions 
\cite{bayer_fine_2002,vamivakas_spin-resolved_2009},
which produce photons but leave the electron spin-state largely unchanged after emission.  These cycling transitions allowed a recent demonstration of single-shot spin-state read-out in the Faraday configuration
\cite{delteil_observation_2014},
but the optical selection rules preclude arbitrary coherent spin manipulation beyond initialization to an eigenstate. Therefore, in order to realize the three essential single-qubit operations of initialization, manipulation, and read-out, there is a need to combine the capabilities of the Voigt and Faraday configurations.

In this Letter, we propose a scheme to read-out the spin-state of a single electron trapped in an optically active quantum dot using a cycling transition induced by the AC Stark effect of a strong optical field far detuned from resonance. In this scheme, a constant Voigt configuration magnetic field allows rapid spin initialization and picosecond spin rotation {\it via} stimulated Raman adiabatic passage 
\cite{imamoglu_quantum_1999,berezovsky_picosecond_2008,press_complete_2008,carter_quantum_2013}.
We show below that the AC Stark effect is capable of modifying the allowed optical transitions resulting in spin-selective cycling transitions, which could be used for a single-shot measurement. 

Here, the AC Stark effect is induced by a circularly-polarized laser that is far
% REMOVED: "red-"
detuned from the optical transitions. In the limit of large detuning, the interaction between the laser and the QD does not significantly populate the excited states, but still shifts the energy levels coupled by that transition 
\cite{unold_optical_2004,xu_fast_2007,muller_emission_2008,muller_creating_2009}.
When this AC Stark shift is much larger than the Zeeman splitting from the Voigt magnetic field, then the polarization selection rules are more similar to the Faraday configuration than the Voigt configuration.  Therefore, we call the combined Voigt field with AC Stark shift the ``pseudo-Faraday'' configuration.  Because the pseudo-Faraday configuration is induced by an optical field rather than a DC magnetic field, the system may be rapidly switched between Voigt and pseudo-Faraday configurations.  This versatility will allow not only spin initialization and manipulation, but also single-shot spin read-out.  We develop a model to determine the time evolution of the density matrix under resonant excitation in the pseudo-Faraday configuration, and use it to demonstrate the feasibility of a single-shot read-out of the electron spin state.

\section{\label{DescriptionOfSystem}Description of the System}

\subsection{\label{Hamiltonian}Hamiltonian}

\begin{figure*}[t]
\includegraphics{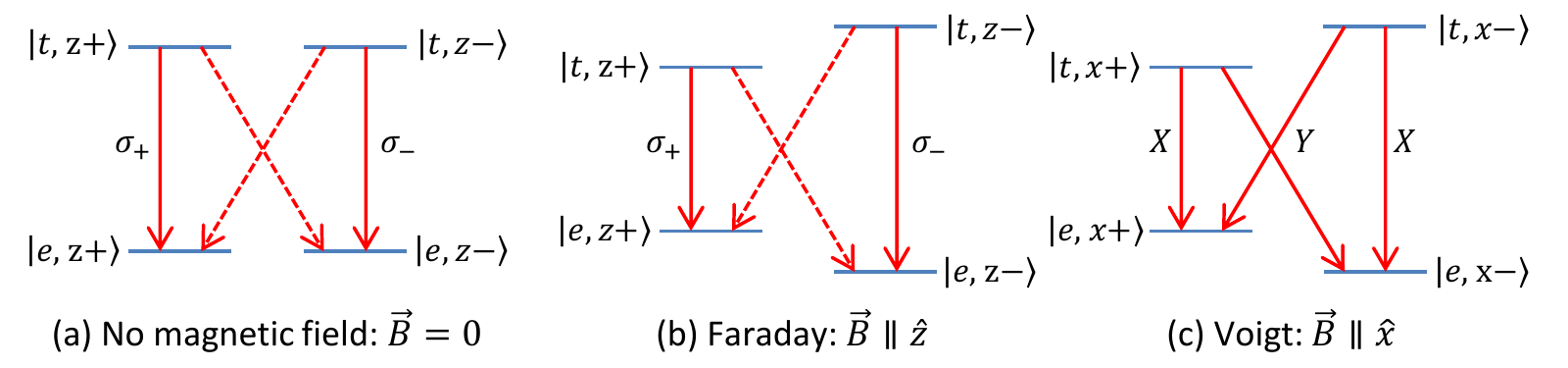}
\caption{\label{fig:Figure1} (Color online) Energy level structure of a charged quantum dot with (a) no magnetic field, (b) a Faraday magnetic field, and (c) a Voigt magnetic field.  Allowed transitions are shown with solid lines; weakly-allowed transitions are shown with dashed lines.}
\end{figure*}

The ground state of the QD can be the empty state, where the lowest lying conduction and highest lying valence bands are empty of electrons and holes respectively. Or the ground state can contain a single charge, either an electron or hole. 
If the QD is in a diode structure, the QD charge can be stabilized and adjusted
depending on the bias voltage applied to the diode contacts 
\cite{warburton_optical_2000}.
In the case of an \textit{n-i}-Schottky diode structure, the QD can have a single electron trapped in the bound conduction band state.  It is the spin-state of this trapped electron that may serve as a qubit
\cite{warburton_single_2013}.
The single-electron $z$-projection spin states are optically coupled to charged exciton (trion) spin states comprising a pair of electrons in the conduction band and a single heavy-hole in the valence band 
\cite{warburton_charged_1997}.
Because the electrons form a singlet state, the spin of the trion is determined solely by the spin of the hole.  One of the electrons may recombine with the hole, emitting a photon and returning the QD to the single electron ground state. We name the relevant eigenstates of the negatively charged QD as
$\ket{e, z+}$, $\ket{e,z-}$, $\ket{t,z+}$, $\ket{t,z-}$
where $e$ means the single electron state, $t$ means the trion state, and $z\pm$  is for the $z$-projection of the spin.  Due to conservation of angular momentum, each trion state has an allowed transition only to the electron state of matching spin, and the transition to the opposite-spin electron state is only weakly allowed due to slight light-hole/heavy-hole mixing 
\cite{calarco_spin-based_2003,atature_quantum-dot_2006,dreiser_optical_2008}.
The energy level structure of the charged QD is shown schematically in Fig.~\ref{fig:Figure1}(a).

The Hamiltonian of a negatively-charged QD in both a magnetic field and an electric field is:
\begin{equation}
H = \hbar \omega_0 (\sigma_{+}^{\dagger} \sigma_{+} + \sigma_{-}^{\dagger} \sigma_{-}) 
	- \vec{\mu} \cdot \vec{B} - \vec{d} \cdot \vec{E}
\label{H}
\end{equation}
where $\omega_0$ is the transition frequency, $\sigma_{+}$ and $\sigma_{-}$ are the lowering operators for the $z$-projection spin-up and spin-down manifolds, respectively, $\vec{\mu}$ and $\vec{d}$ are the magnetic and electric dipole operators, and $\vec{B}$ and $\vec{E}$ are the magnetic and electric field amplitudes.  The lowering operators are defined in terms of the electron and trion spin-states as
\begin{equation} \label{eq:QD_lowering_operators}
\begin{array}{l}
\sigma_{+} = \ket{e,z+} \bra{t,z+}	\\
\sigma_{-} = \ket{e,z-} \bra{t,z-}
\end{array}
\end{equation}
We decompose $H$ into ``atomic'' ($H_{A}$), magnetic dipole ($H_{Z}$), and electric dipole ($H_{D}$) components ($H=H_{A}+H_{Z} + H_{D}$). With no magnetic field, the $z\pm$  spin projection states are degenerate.  A Voigt configuration magnetic field is perpendicular to the propagation direction of the emitted light, which is typically the $z$-direction, normal to the sample surface.  In that case $\vec{B}=\hat{x}B_{x}$, and the magnetic dipole, or Zeeman, Hamiltonian becomes
\cite{van_kesteren_fine_1990}
\begin{equation}
H_Z = \mu_B B_x ( g_{e,x} (s_e^{\dagger} + s_e) + g_{h,x} (s_h^{\dagger} + s_h) )
\end{equation}
where $\mu_B$ is the Bohr magneton, $g_{e,x}$ and $g_{h,x}$ are the electron and hole g-factors for a magnetic field in the $x$-direction, and $s_e$ and $s_h$ are the electron and hole spin-flip operators, respectively.  The spin-flip operators couple states of similar charge configuration but opposite spin.  In terms of the electron and trion spin-states, the spin-flip operators are defined as
\begin{equation}
\begin{array}{l}
s_e = \ket{e,z-} \bra{e,z+}	\\
s_h = \ket{t,z-} \bra{t,z+}
\end{array}
\end{equation}
The form of the Zeeman Hamiltonian in the Voigt configuration leads to eigenstates that are superpositions of the zero-field spin-states.

To describe the AC Stark shift, we assume an oscillatory form for the electric field, as in a single-frequency laser beam, and perform the standard rotating wave approximation 
\cite{scully_quantum_1997}
to obtain a Hamiltonian for the unperturbed QD and the electric dipole interaction:
\begin{eqnarray} \label{eq:Hdipole}
H_A + H_D = && \frac{1}{2} \{ \hbar \Delta_1 (\sigma_{+}^{\dagger} \sigma_{+} - \sigma_{+} \sigma_{+}^{\dagger}
								  +\sigma_{-}^{\dagger} \sigma_{-} - \sigma_{-} \sigma_{-}^{\dagger}) \nonumber \\
					&&	+ d (E_{1+} \sigma_{+} + E_{1-} \sigma_{-} + H.c.) \}
\end{eqnarray}
where $\Delta_1 = \omega_0 - \omega_1$ is the detuning of the laser frequency $\omega_1$ from the QD resonance $\omega_0$; $d$ is the dipole moment of the transitions; $E_{1+}$ and $E_{1-}$ are the complex amplitudes of the left and right circularly polarized components of the electric field; and $H.c.$ means Hermitian conjugate.  We can further simplify this expression by introducing the complex Rabi frequencies associated with the two circularly polarized components: $\Omega_{1+}=d E_{1+} / \hbar$ and $\Omega_{1-}=d E_{1-} / \hbar$. We are primarily interested in circularly-polarized light because it will produce an AC Stark shift that reduces the effect of the magnetic coupling between the spin-states, as we will demonstrate below. Therefore, we henceforth assume left-circularly polarized light $(\Omega_{1-}=0)$, and without loss of generality we can treat $\Omega_{1+}$ as purely real.

Using $H_{A}$, $H_{Z}$ and $H_{D}$ from above, we can now write the Hamiltonian from Eqn.~(\ref{H}) of the charged QD system in the rotating frame as
\begin{eqnarray} \label{eq:H0}
H_0 = && \frac{1}{2} \hbar \Delta_1 (\sigma_{+}^{\dagger} \sigma_{+} - \sigma_{+} \sigma_{+}^{\dagger}
					  +\sigma_{-}^{\dagger} \sigma_{-} - \sigma_{-} \sigma_{-}^{\dagger}) \nonumber \\
	&& + \mu_B B_x ( g_{e,x} (s_e^{\dagger} + s_e) + g_{h,x} (s_h^{\dagger} + s_h) ) \nonumber \\
	&& + \frac{1}{2} \hbar \Omega_{1+} (\sigma_{+}^{\dagger} + \sigma_{+})
\end{eqnarray}
In the basis of the unperturbed QD eigenstates in the rotating frame, the matrix representation of the Hamiltonian is
\begin{equation} \label{eq:H0matrix}
H_0 = \left[
\begin{matrix}
-\hbar \Delta_1 / 2 	& \mu_B B_x g_{e,x} 	& \hbar \Omega_{1+} / 2 & 0 \\
\mu_B B_x g_{e,x} 		& -\hbar \Delta_1 / 2 	& 0 					& 0 \\
\hbar \Omega_{1+} / 2 	& 0 					& \hbar \Delta_1 / 2 	& -\mu_B B_x g_{h,x}\\
0 						& 0 					& -\mu_B B_x g_{h,x} 	& \hbar \Delta_1 / 2\\
\end{matrix}
\right]
\end{equation}
and the unperturbed eigenstates are represented as vectors
\begin{equation} \label{eq:FaradayBasis}
\begin{matrix}
\ket{e,z+}= \left[
\begin{matrix}
1 \\ 0 \\ 0 \\ 0
\end{matrix}
\right]
&
\ket{e,z-}= \left[
\begin{matrix}
0 \\ 1 \\ 0 \\ 0
\end{matrix}
\right]
\\
\ket{t,z+}= \left[
\begin{matrix}
0 \\ 0 \\ 1 \\ 0
\end{matrix}
\right]
&
\ket{t,z-}= \left[
\begin{matrix}
0 \\ 0 \\ 0 \\ 1
\end{matrix}
\right]
&
\end{matrix}
\end{equation}
The interaction represented by $\Omega_{1+}$ and $\Delta_1$ is that of the far 
%"red-"
detuned laser that will cause the AC Stark effect.

\subsection{\label{Zeeman}Zeeman Effect}
In the presence of a magnetic field in the Voigt geometry with $B_x \ge 0$ (and no laser field), the eigenstates are no longer the $z\pm$  projections of the spin, but the $x\pm$  projections, which are superpositions of the $z$-projection states.  We can diagonalize the system Hamiltonian $H_0$ with $\Omega_{1+}=0$ and $\Delta_1 = 0$ to obtain the eigenvalues
\begin{eqnarray}
\lambda_1 = && -\mu_B B_x g_{e,x} \nonumber \\
\lambda_2 = && \mu_B B_x g_{e,x}  \nonumber \\
\lambda_3 = && -\mu_B B_x g_{h,x} \nonumber \\
\lambda_4 = && \mu_B B_x g_{h,x}  \nonumber
\end{eqnarray}
and eigenstates
\begin{equation}
\begin{matrix}
\ket{e,x-} = \frac{1}{\sqrt{2}} \left[
\begin{matrix}
1 \\ -1 \\ 0 \\ 0
\end{matrix}
\right]
&
\ket{e,x+} = \frac{1}{\sqrt{2}} \left[
\begin{matrix}
1 \\ 1 \\ 0 \\ 0
\end{matrix}
\right]
\nonumber \\
\ket{t,x+} = \frac{1}{\sqrt{2}} \left[
\begin{matrix}
0 \\ 0 \\ 1 \\ 1
\end{matrix}
\right]
&
\ket{t,x-} = \frac{1}{\sqrt{2}} \left[
\begin{matrix}
0 \\ 0 \\ 1 \\ -1
\end{matrix}
\right]
&
\end{matrix}
\end{equation}
These representations of the eigenstates are in the $z\pm$ basis, thus we can see that the $x\pm$ projections are superpositions of the $z$-projection states.  The left side of Fig.~\ref{fig:Figure2} shows the Zeeman splitting of the electron and trion energy levels for a Voigt geometry field. In a Voigt configuration, transitions from either trion spin-state to either electron spin-state are allowed, as depicted schematically in Fig. \ref{fig:Figure1}(c).  There are no cycling transitions that might allow a single-shot fluorescence measurement of the electron eigenstate.

\subsection{\label{ACStark}AC Stark Effect}
In the absence of a magnetic field and in the large detuning limit, the energy levels coupled by the electric dipole interaction are modified by the AC Stark shift 
\cite{cohen-tannoudji_theorie_1962,dupont-roc_lifting_1967,bonch-bruevich_current_1968,cohen-tannoudji_experimental_1972}.
We can see this by determining the eigenvalues of the Hamiltonian $H_0$ from Eqn.~(\ref{eq:H0matrix}) with $B_x=0$:
\begin{eqnarray}
\lambda_1 = && -\frac{\hbar}{2} \mathcal{W}_1 \nonumber \\
\lambda_2 = && -\frac{\hbar}{2} \Delta_1      \nonumber \\
\lambda_3 = && \frac{\hbar}{2} \mathcal{W}_1  \nonumber \\
\lambda_4 = && \frac{\hbar}{2} \Delta_1       \nonumber
\end{eqnarray}
where $\mathcal{W}_1 = \sqrt{\Delta_1^2 + \Omega_{1+}^2}$ is the generalized Rabi frequency for the far-detuned laser. The eigenvectors of the Hamiltonian, still expressed in the basis of the unperturbed states, are
\begin{equation}
\begin{matrix}
\vec{v}_1 = \frac{1}{\sqrt{2} \sqrt{\mathcal{W}_1^2 - \mathcal{W}_1 \Delta_1}} \left[
\begin{matrix}
\Omega_{1+} \\ 0 \\ \Delta_1 - \mathcal{W}_1 \\ 0
\end{matrix}
\right]
& \qquad
\vec{v}_2 = \left[
\begin{matrix}
0 \\ 1 \\ 0 \\ 0
\end{matrix}
\right]
\nonumber \\
\vec{v}_3 = \frac{1}{\sqrt{2} \sqrt{\mathcal{W}_1^2 + \mathcal{W}_1 \Delta_1}} \left[
\begin{matrix}
\Omega_{1+} \\ 0 \\ \Delta_1 + \mathcal{W}_1 \\ 0
\end{matrix}
\right]
& \qquad
\vec{v}_4 = \left[
\begin{matrix}
0 \\ 1 \\ 0 \\ 0
\end{matrix}
\right]
&
\end{matrix}
\end{equation}
\begin{figure}[t]
\includegraphics{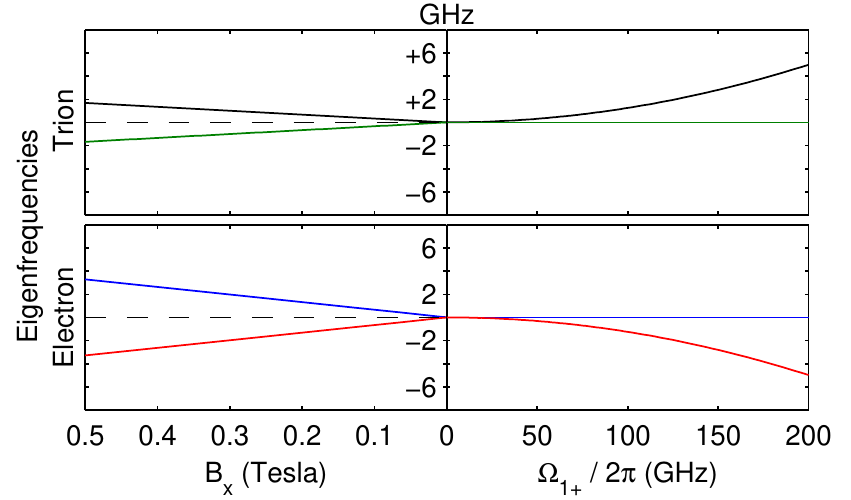}
\caption{\label{fig:Figure2} (Color online) Splitting of the electron and trion spin states due to the Zeeman effect (left) and the AC Stark effect (right).  The trion eigenfrequencies are plotted relative to the zero-field transition resonance frequency $\omega_0 / 2\pi$ (dashed), for a fixed laser detuning of $\Delta_{1}/2\pi = 2000$ GHz.}
\end{figure}

The eigenstates corresponding to the $z\pm$ manifold are no longer purely electronic or trionic, but a superposition of both.  This occurs because they are coupled by the $\sigma_+$ polarization of the laser.  If we make the assumption that the detuning $\Delta_1$ is much larger than the Rabi frequency $\Omega_{1+}$, then to first order in the ratio $\Omega_{1+} / \Delta_1$ the generalized Rabi frequency is
\begin{equation}
\mathcal{W}_1 \approx \Delta_1 + \frac{\Omega_{1+}^2}{2 \Delta_1}
\end{equation}
and we can approximate the eigenvalues and eigenvectors as follows.
\begin{equation} \label{eq:ACStarkEigensystem}
\begin{array}{l}
\lambda_1 = - \frac{\hbar}{2} \Delta_1 - \frac{\hbar\Omega_{1+}^2}{4\Delta_1} 
			+ \mathcal{O}(\frac{\Omega_{1+}^2}{\Delta_1^2})
\\
\lambda_3 = \frac{\hbar}{2} \Delta_1 + \frac{\hbar\Omega_{1+}^2}{4\Delta_1} 
			+ \mathcal{O}(\frac{\Omega_{1+}^2}{\Delta_1^2})
\\
\vec{v}_1 = \left[
\begin{matrix}
1 \\ 0 \\ -\Omega_{1+} / 2\Delta_1 \\ 0
\end{matrix}
\right] + \mathcal{O}(\Omega_{1+}^2 / \Delta_1^2)
\\
\vec{v}_3 = \left[
\begin{matrix}
\Omega_{1+} / 2\Delta_1 \\ 0 \\ 1 \\ 0
\end{matrix}
\right] + \mathcal{O}(\Omega_{1+}^2 / \Delta_1^2)
\end{array}
\end{equation}

Note that these eigenvectors are normalized only to first order in $\Omega_{1+} / \Delta_1$.  In the large detuning approximation, one state, $\vec{v}_1$, is more electron-like and the other, $\vec{v}_3$, is more trion-like.  
% REMOVED: Having made the rotating wave approximation, the eigenvectors are expressed in the rotating frame.
% ADDED:
Because the Hamiltonians in Eqns.~(\ref{eq:Hdipole}), (\ref{eq:H0}), and (\ref{eq:H0matrix}) are expressed in the rotating frame in order to make the rotating wave approximation, the eigenvectors in Eqn.~(\ref{eq:ACStarkEigensystem}) are also expressed in the rotating frame.  
Therefore, to determine the energies of the states we must add $\hbar(\omega_0 + \omega_1) / 2$ to the eigenvalues for the trion-like states, $\lambda_3$ and $\lambda_4$, and add $\hbar(\omega_0 - \omega_1) / 2$ to those for the electron-like states, $\lambda_1$ and $\lambda_2$:
\begin{eqnarray} \label{eq:ACStarkEnergies}
\begin{array}{ll}
E_1 \approx - \frac{\hbar\omega_{1+}^2}{4\Delta_1}
&
E_2 = 0
\\
E_3 \approx \hbar \omega_0 + \frac{\hbar\omega_{1+}^2}{4\Delta_1}
&
E_4 = \hbar \omega_0
\end{array}
\end{eqnarray}
We can see from the above expressions that a circularly-polarized far-detuned laser results in a spin-selective AC Stark shift from the unperturbed energies.  For the case of a $\sigma_+$ polarized field that we consider here, the $z+$  manifold states are shifted ($E_1$ \& $E_3$), while the $z-$ manifold states are not ($E_2$ \& $E_4$).  For red-detuning ($\Delta_1>0$), the electron $z+$  energy shifts downward and the trion $z+$  energy shifts upward by the same amount.
% ADDED:
For blue-detuning ($\Delta_1<0$), the energy shifts would be the opposite.

% ADDED:
The purpose of the AC Stark laser is to shift the energies without populating the trion-like states.  A red-detuned laser is preferred over a blue-detuned one because of the lower probability of inelastic absorption for a red-detuned laser.  Inelastic absorption is the absorption of a photon combined with either emission or absorption of a phonon from the crystal lattice.  Inelastic absorption of a red-detuned photon would require the simultaneous absorption of a phonon, which is improbable due to the low temperature at which optical experiments on QDs are usually performed.  Inelastic absorption of a blue-detuned photon, however, would require only the emission of a phonon, and that process can still occur even at zero temperature.  Therefore, a red-detuned AC Stark laser causes the energy states to shift while minimizing the probability of exciting the QD.  Subsequently we will assume that the AC Stark laser is red-detuned.

The amount of state mixing in the $z+$  manifold is proportional to $\Omega_{1+}/\Delta_1$, and can thus be reduced arbitrarily by increasing the detuning.  The energy shift, however is proportional to $\Omega_{1+}^2/\Delta_1$.  Thus, with enough laser power we can have a situation where the state mixing can be made negligible while maintaining a non-zero energy shift.  The right side of Fig.~\ref{fig:Figure2} shows the shifting of the energy levels as a function of Rabi frequency $\Omega_{1+}$ at a fixed detuning of $\Delta_1/2\pi=2000$ GHz.  At fixed large detuning, the energy shift is quadratic in $\Omega_{1+}$, and for $\sigma_{+}$ polarization it only affects the $z+$ manifold states.

\subsection{\label{PseudoFaraday}Pseudo-Faraday Configuration}
The Voigt geometry magnetic field couples the $z+$  spin states to the $z-$ spin states, while the AC Stark effect of a $\sigma_+$ polarized laser shifts only the $z+$  spin states.  In the presence of both a magnetic field and a strong, far red-detuned $\sigma_+$ polarized laser, the spin projection of the eigenstates and the polarization selection rules result from a competition between the magnetic coupling between the $z+$  and $z-$ states and the AC Stark shift of the $z+$  states.  When the AC Stark shift is very large compared to the magnetic coupling, then the magnetic field has little effect on the system.  We call this situation the pseudo-Faraday configuration because the energy structure, eigenstates, and polarization selection rules are similar to the Faraday magnetic field configuration where $\vec{B} = \hat{z} B_z$.  We can calculate the energy levels from the eigenvalues of the Hamiltonian $H_0$ from Eqn.~(\ref{eq:H0matrix}), but because that representation is in the rotating frame, to obtain the eigenenergies we must perform the same operation as for Eqns.~(\ref{eq:ACStarkEnergies}), above.
Because the pseudo-Faraday eigenvalues are roots of the 4th-order characteristic polynomial of $H_0$ they are very complicated, thus we forgo including them here.  Instead, we calculate them numerically and discuss specific points of interest.

\begin{figure}[t]
	\begin{minipage}{0.71\columnwidth} %0.71
		\includegraphics[width=\columnwidth]{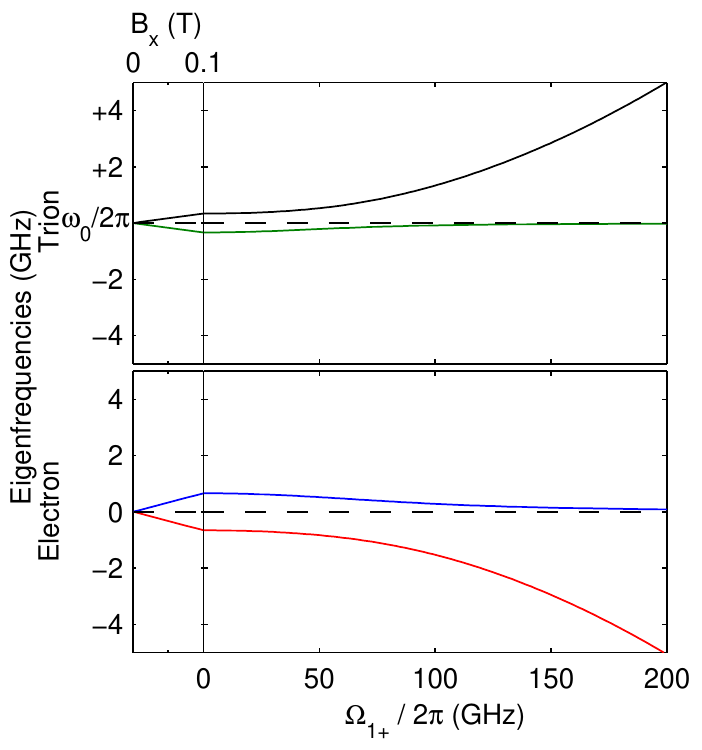}
	\end{minipage}
	\hspace{-0.035\columnwidth}
	\begin{minipage}{0.26\columnwidth} %0.26
		\includegraphics[width=\columnwidth]{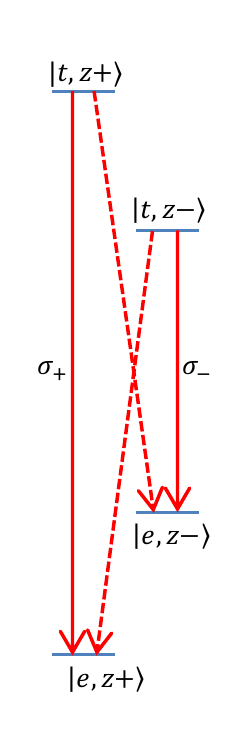}
	\end{minipage}
	\caption{\label{fig:Figure3} (Color online) Evolution of eigenstate frequencies as a function of first magnetic field, and then Rabi frequency with a fixed magnetic field of 0.1 T and detuning  $\Delta_1/2\pi=2000$ GHz.  The diagram to the right shows the energy level structure in the pseudo-Faraday configuration.  Allowed transitions are shown as solid lines and labeled with their polarizations.  Weakly-allowed transitions are shown as dashed lines.}
\end{figure}

Figure \ref{fig:Figure3} depicts the evolution of the QD eigenstate frequencies ($\propto$ energy) for finite magnetic field and large detuning, $\Delta_1 \gg \Omega_{1+}$. In the first section (left of the vertical  line) $\Omega_{1+}=0$ and the magnetic field increases, causing the previously degenerate electron and trion levels to split and the eigenstates to become the $x\pm$  spin projections, which are superpositions of the $z\pm$  spin states.  In this configuration, transitions from either trion spin-state to either electron spin-state are allowed. In the second section (right of the vertical line), the magnetic field is fixed at 0.1 T and $\Omega_{1+}$ is increased. The AC Stark effect shifts the $z+$  components of the electron and trion states, but not the $z-$ components. When the AC Stark shift becomes much larger than the Zeeman splitting, the eigenstates are more like those of the Faraday configuration, except that the $z-$ manifold is unperturbed. The final energy-level configuration is shown to the right of the plot; the allowed transitions are circularly-polarized and shown with solid lines, while weakly-allowed transitions are shown with dotted lines.

The true Faraday configuration has cycling transitions that allow a single-shot measurement of the electron spin-state 
\cite{delteil_observation_2014}.
As depicted in Fig.~\ref{fig:Figure1}(b), the two trion states of the QD each have two possible transitions, and the ratio of their probabilities -- or emission rates -- is called the branching ratio.  In a cycling transition, the branching ratio is very different from unity, meaning that the spin-preserving transition is far more likely than the spin-flipping transition.  The branching ratio quantifies the measurement back-action of the cycling transition and will determine the number of cycles that can be used for a single-shot measurement.  

To obtain the branching ratio, we take the ratio of the spin-flipping transition rate to the spin-preserving transition rate for one of the trion-like states:
\begin{equation} \label{eq:branching_ratio}
r_B = \frac{\left| \Braket{ \psi_1 | \mathbf{d} | \psi_4 } \right|^2}
		   {\left| \Braket{ \psi_2 | \mathbf{d} | \psi_4 } \right|^2}
\end{equation}
where $\ket{\psi_i}$ are the pseudo-Faraday eigenstates that are the eigenvectors of the Hamiltonian $H_0$ in Eqn.~(\ref{eq:H0matrix}), and $\mathbf{d}$ is the dipole moment operator (see the appendix for the derivation of Eqn.~(\ref{eq:branching_ratio})).

\begin{figure}[t]
\includegraphics{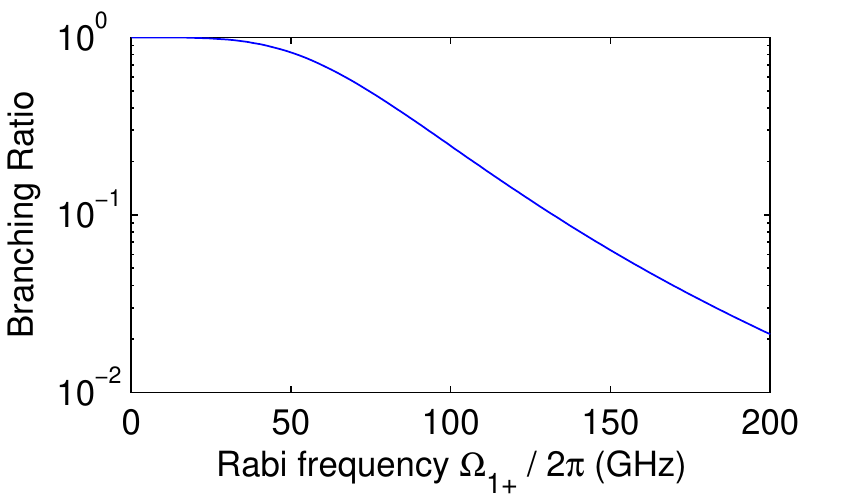}
\caption{\label{fig:Figure4} (Color online) Evolution of the branching ratio from the Voigt configuration ($B_x=0.1$ T) to the pseudo-Faraday configuration.  In the pseudo-Faraday configuration, the branching ratio is 0.02.}
\end{figure}

In Fig.~\ref{fig:Figure4}, we plot the branching ratio as a function of Rabi frequency for fixed magnetic field and typical g-factors
\cite{kroner_resonant_2008}:
$g_{e,x}=0.47$, $g_{h,x}=0.24$.
As the Rabi frequency increases, the branching ratio reduces significantly, becoming similar to that of the Faraday configuration.  For a magnetic field of 0.1 T, detuning of $\Delta_1/2\pi = 2$ THz, and Rabi frequency of $\Omega_{1+}/2\pi=200$ GHz, we calculate the branching ratio to be 0.02.  Measured values of branching ratio for solid-state systems range from 0.001 
\cite{fernandez_optically_2009}
to 0.04 
\cite{muller_optical_2014}.
Our prediction of the branching ratio for a QD in the pseudo-Faraday configuration is within this range, suggesting that it may be possible to perform a single-shot read-out of the electron spin-state {\it via} resonantly excited fluorescence.

\section{\label{SpinReadOut}Spin Read-out Operation}

\subsection{\label{Overview}Overview}
We have demonstrated that applying a strong, far red-detuned, circularly-polarized laser to a charged QD in a Voigt configuration magnetic field results in a situation similar to a Faraday magnetic field.  We now discuss how a practical spin read-out scheme would work in the pseudo-Faraday configuration.  To perform a spin read-out operation, we need to excite the system in a spin-selective manner and detect the fluorescence.  Detection of a photon would correspond to the electron being in a certain spin state, and the fidelity of the measurement depends on the branching ratio and the spin-selectivity of the excitation.

In the pseudo-Faraday configuration, there are two spin-preserving transitions that are non-degenerate and two weakly-allowed spin-flipping transitions that are nearly degenerate (see Fig.~\ref{fig:Figure3}).  The spin-preserving transitions are circularly-polarized, $\sigma_{+}$ and $\sigma_{-}$, and can both be excited by linearly-polarized light.  The energy difference between the two spin-preserving transitions, however, makes resonant excitation of one of them a spin-selective excitation even when linear polarization is used.  Thus, spin-selective excitation may be accomplished by a linearly-polarized laser tuned to resonance with one of the spin-preserving transitions.  The fluorescence will be circularly-polarized and may, therefore, be distinguished from the linearly-polarized laser scattering by cross-polarized detection 
\cite{vamivakas_spin-resolved_2009}.
Alternatively, we can use a modal discrimination method, as in 
Refs.~[\onlinecite{muller_resonance_2007}] and [\onlinecite{flagg_resonantly_2009}],
wherein the resonant laser is introduced into the waveguide mode of a planar microcavity that confines the laser scattering, while the QD fluorescence is emitted into the orthogonal cavity mode.

After the electron has undergone some operations (e.g. initialization and/or manipulation) in the Voigt configuration, the AC Stark laser can be applied to transform to the pseudo-Faraday configuration where read-out will occur.
The application of the AC Stark laser can be rapid compared to the switching time of a magnetic field, but if the laser field is applied too fast then the transition from Voigt to pseudo-Faraday will be non-adiabatic.
Prior to the application of the AC Stark laser, the electron may be in any arbitrary superposition of the two Voigt eigenstates, which are the $x\pm$  projection states of the spin.  The transition from Voigt to pseudo-Faraday configurations will occur adiabatically if the AC Stark laser is turned on slowly relative to $\hbar/\delta_e$, where $\delta_e = 2 \mu_B g_{e,x} B_x$ is the energy splitting of the electron spin states in the Voigt configuration.  For a 0.1 T magnetic field and typical electron g-factor
\cite{kroner_resonant_2008}
$g_{e,x}=0.24$, the electron spin precession period is $\hbar/\delta_e = 120$ ps.  Therefore, the AC Stark laser must have a rise time of about 1 ns or greater.  If the adiabatic condition is satisfied, the population of the $x\pm$  state in the Voigt configuration transitions without change to the population of the $z\mp$  state in the pseudo-Faraday configuration.  This is for the case of a $\sigma_{+}$ polarized AC Stark laser; for a $\sigma_{-}$ polarized laser, the mapping from $x$-basis to $z$-basis would be the opposite.  Due to the adiabatic mapping of the Voigt $x$-basis to the pseudo-Faraday eigenbasis, detection of the spin-selectively excited fluorescence in the pseudo-Faraday configuration is equivalent to a projective measurement of the spin onto the $x\pm$ states in the Voigt configuration.
Furthermore, a measurement in any arbitrary spin projection basis can be performed by preceding the AC Stark laser with a rotation of the electron spin Bloch sphere {\it via} stimulated Raman adiabatic passage by additional laser pulses
\cite{berezovsky_picosecond_2008,press_complete_2008}.
The capability of using multiple measurement bases for identically prepared states would allow full quantum state tomography 
\cite{james_measurement_2001}
to be performed on the electron spin state.

\subsection{\label{FLSM}Floquet-Liouville Supermatrix Approach}
To demonstrate the feasibility of a single-shot read-out of the electron spin state we numerically calculate the time evolution of the QD density matrix $\rho(t)$ under the conditions outlined in the previous section.  This requires the addition of another electric dipole Hamiltonian describing a second, near-resonant laser at a frequency $\omega_2$ that can spin-selectively excite population from the electron states to the trion states.  In the rotating frame with rotation frequency $\omega_1$, the Hamiltonian of the second laser interaction is oscillatory:
\begin{equation} \label{eq:Hsecondlaser}
H_1 = \frac{\hbar}{2}
\left[
\begin{matrix}
0				&	0				&	0	&	0 \\
0				&	0				&	0	&	0 \\
\Omega_{2+}^*	&	0				&	0	&	0 \\
0				&	\Omega_{2-}^*	&	0	&	0
\end{matrix}
\right]
e^{i(\Delta_2 - \Delta_1)t} + H.c.
\end{equation}
where $\Delta_2 = \omega_0 - \omega_2$ is the detuning of the near-resonant laser, and $\Omega_{2+} = d E_{2+} / \hbar$ and $\Omega_{2-} = d E_{2-} / \hbar$ are the complex Rabi frequencies for the two circularly polarized components of the electric field.  The total Hamiltonian of the charged QD system, including magnetic and electric field interactions, can be expressed as
\begin{equation} \label{eq:Hdecomp}
H(t) = H^{(0)} + H^{(1)} e^{i(\Delta_2 - \Delta_1)t} + H^{(-1)} e^{-i(\Delta_2 - \Delta_1)t}
\end{equation}
where $H^{(0)}$ is the time-independent $H_0$ from Eqn.~(\ref{eq:H0matrix}) and $H^{(\pm 1)}$ are the two constant matrices in Eqn.~(\ref{eq:Hsecondlaser}).

The time evolution of the density matrix can be determined by solving the Liouville equation, which can be extended to include spontaneous transitions by using a Lindblad superoperator $\mathcal{L}(\rho)$ 
\cite{lindblad_generators_1976}:
\begin{equation} \label{eq:Liouville}
\frac{\partial}{\partial t} \rho(t) = - \frac{i}{\hbar} \left[H(t), \rho(t)\right] + \mathcal{L}(\rho)
\end{equation}
In solving the Liouville equation numerically 
% ADDED:
using a standard differential equation solving algorithm, 
the oscillatory nature of the Hamiltonian $H(t)$ requires the integration time-step to be
% ADDED:
much
smaller than the oscillation period $2\pi / (\Delta_2 - \Delta_1)$.  The interesting system dynamics, however, occur on a time scale much longer than the oscillation period.
% REMOVED: This makes the Liouville equation inefficient to solve numerically in this form.  
% ADDED:
Thus it is inefficient to use this form of the Liouville equation to numerically solve for the long-term dynamics of the system.
Instead, we solve the equation using a Floquet-Liouville supermatrix approach 
\cite{ho_floquet-liouville_1986},
which we describe here for the specific case of interest.  Similar Floquet theory approaches have been used to describe the spectrum of resonance fluorescence from QDs under bichromatic near-resonant excitation 
\cite{agarwal_spectrum_1991,ficek_resonance-fluorescence_1993,aronstein_comment_2002,peiris_bichromatic_2014}.
Here, however, we need not the emission spectrum but the density matrix evolution.
% ADDED:
The Floquet-Liouville supermatrix approach allows analytical solutions to any order of approximation, meaning that the density matrix at any time can be calculated without needing to calculate all the intervening density matrix values.

The unperturbed electron and trion eigenstates of $H_0$ form a complete orthonormal basis $\{ \ket{\alpha} \}$ in the Hilbert space of the QD states.  In this basis, the density matrix operator can be expressed using the matrix elements $\rho_{\alpha \beta} (t) \equiv \braket{\alpha|\rho(t)|\beta}$:
\begin{equation}
\rho(t) = \sum\limits_{\alpha \beta} \rho_{\alpha \beta} (t) \ket{\alpha} \bra{\beta}
\end{equation}
This expression is often considered a matrix in the state basis $\{ \ket{\alpha} \}$, but it can also be considered a supervector $\vec{\rho}(t)$ in the operator basis $\{ \ket{\alpha} \bra{\beta} \}$ with elements $\rho_{\alpha \beta} (t)$.  This interpretation unfolds the 4-by-4 density matrix into a 16-dimensional density supervector.  A similar transformation can be performed on the Liouville equation, which is a 4-by-4 matrix equation in the Hilbert space of the QD states.  Utilizing the completeness of the $\{ \ket{\alpha} \}$ basis, 
\begin{equation}
1 = \sum\limits_{\alpha} \ket{\alpha} \bra{\alpha}
\end{equation}
we can express Eqn.~(\ref{eq:Liouville}) as
\begin{widetext}
\begin{equation} \label{eq:LiouvilleIndexForm}
\frac{\partial}{\partial t} \sum\limits_{\alpha\beta} \rho_{\alpha\beta} \ket{\alpha} \bra{\beta}
= \sum\limits_{\alpha\beta} \left(
- \frac{i}{\hbar} \sum\limits_{k} \left( H_{\alpha k} \rho_{k \beta} - \rho_{\alpha k} H_{k \beta} \right)
+ \left( \mathcal{L}(\rho) \right)_{\alpha\beta}
\right)
\ket{\alpha} \bra{\beta}
\end{equation}
\end{widetext}
For clarity we have dropped the explicit time dependence of $H(t)$ and $\rho(t)$.  While the Hamiltonian can be expressed as a 4-by-4 matrix operating on the density matrix, as in the above equation, the Lindblad superoperator cannot.  $\mathcal{L}(\rho)$ represents population relaxation processes (e.g., spontaneous emission) and decoherence processes (e.g., pure dephasing and all population relaxation). Population relaxation, or $T_1$ processes, are the $\alpha=\beta$ elements of $(\mathcal{L}(\rho))_{\alpha\beta}$:
\begin{equation} \label{eq:poprelax}
(\mathcal{L}(\rho))_{\alpha\alpha} = \sum\limits_{q} \left(
-\Gamma_{\alpha q} \rho_{\alpha \alpha} + \Gamma_{q \alpha} \rho_{q q}
\right)
\end{equation}
where $\Gamma_{\alpha \beta}$ is the spontaneous transition rate from state $\ket{\alpha}$ to state $\ket{\beta}$. The first term in Eqn.~(\ref{eq:poprelax}) represents transitions from state $\ket{\alpha}$ to all the other states; the second term is transitions to state $\ket{\alpha}$ from all the other states.  The population relaxation rates form a 4-by-4 matrix with elements $\Gamma_{\alpha \beta}$ that is not generally symmetric: transitions from trion states to electron states occur spontaneously, but not the reverse.  Examining Eqn.~(\ref{eq:poprelax}), the effects of the diagonal elements $\Gamma_{\alpha \alpha}$ cancel out, and therefore without loss of generality we can set them all to zero.  Decoherence, or $T_2$ processes, are the $\alpha\neq\beta$ elements of $(\mathcal{L}(\rho))_{\alpha\beta}$:
\begin{equation} \label{eq:decoherence}
(\mathcal{L}(\rho))_{\alpha\beta} = \left(
- \frac{1}{2} \sum\limits_{q}
\left( \Gamma_{\alpha q} + \Gamma_{\beta q} \right) - \gamma_{\alpha \beta}
\right) \rho_{\alpha \beta}
\end{equation}
where $\gamma_{\alpha\beta}$ is the pure dephasing rate for the coherence $\rho_{\alpha\beta}$ between the states $\ket{\alpha}$ and $\ket{\beta}$.  The terms in the sum are the decoherence caused by population relaxation.  The second term is pure dephasing, also called homogeneous broadening.  The pure dephasing rates form a 4-by-4 matrix with elements $\gamma_{\alpha\beta}$ that is symmetric: $\gamma_{\alpha\beta}=\gamma_{\beta\alpha}$.

In contrast with the Hamiltonians, the Lindblad superoperator cannot be expressed as a single 4-by-4 matrix operating on the density matrix.  However, with the above decompositions of $\mathcal{L}(\rho)$ and identities such as 
\begin{equation}
\rho_{\alpha\beta} = \sum\limits_{\mu\nu} \delta_{\alpha\mu} \delta_{\beta\nu} \rho_{\mu\nu}
\end{equation}
where $\delta_{ij}$ is the Kronecker delta, we can rearrange Eqn.~(\ref{eq:LiouvilleIndexForm}) into a supermatrix equation that allows the both the commutator with $H(t)$ and the Lindblad superoperator to be expressed as a single 16-by-16 supermatrix operating on $\vec{\rho}(t)$:
\begin{equation} \label{eq:supervectorLiouville}
\frac{\partial}{\partial t} \vec{\rho}(t) = - \frac{i}{\hbar} L(t) \vec{\rho}(t)
\end{equation}
Or, expressed using the supervector and supermatrix elements:
\begin{equation}
\frac{\partial}{\partial t} \rho_{\alpha\beta}(t) = - \frac{i}{\hbar} 
\sum\limits_{\mu\nu} L_{\alpha\beta;\mu\nu}(t) \rho_{\mu\nu}(t)
\end{equation}
The elements of the Liouville supermatrix $L(t)$ are:
\begin{widetext}
\begin{equation}
\begin{array}{ll}
(\alpha=\beta)
&
\qquad L_{\alpha\alpha;\mu\nu}(t) = 
\left( H_{\alpha\mu}(t) \delta_{\alpha\nu} - H_{\nu\alpha}(t) \delta_{\alpha\mu} \right)
+ i\hbar \left( \Gamma_{\mu\alpha} \delta_{\mu\nu} - \sum\limits_{q} \Gamma_{\alpha q}\delta_{\alpha\mu}\delta_{\alpha\nu} \right)
\\
(\alpha\neq\beta)
&
\qquad L_{\alpha\beta;\mu\nu}(t) =
\left( H_{\alpha\mu}(t) \delta_{\beta\nu} - H_{\nu\beta}(t) \delta_{\alpha\mu} \right)
+ i\hbar \left( - \frac{1}{2} \sum\limits_{q} \left( \Gamma_{\alpha q} + \Gamma_{\beta q} \right) 
- \gamma_{\alpha\beta} \right) \delta_{\alpha\mu} \delta_{\beta\nu}
\end{array}
\end{equation}
\end{widetext}

Although in Eqn.~(\ref{eq:supervectorLiouville}) the Liouville equation is now expressed as an ordinary differential equation with a single matrix, the matrix still has an oscillatory time-dependence. Similar to how the Hamiltonian is separated in Eqn.~(\ref{eq:Hdecomp}), we can separate $L(t)$ into a constant term and two oscillatory terms:
\begin{equation} \label{eq:Lexpansion}
L(t) = L^{(0)} + L^{(1)} e^{i(\Delta_2-\Delta_1)t} + L^{(-1)} e^{-i(\Delta_2-\Delta_1)t}
\end{equation}
where $L^{(0)}$ contain the constant part of $H(t)$ and all of the relaxation terms from $\mathcal{L}(\rho)$, and $L^{(\pm 1)}$ contains only the Hamiltonians $H^{(\pm 1)}$.  In detail, the supermatrix terms in $L(t)$ are:
\begin{widetext}
\begin{equation}
\begin{array}{ll}
& \qquad
L^{(\pm 1)}_{\alpha\beta;\mu\nu} = H^{(\pm 1)}_{\alpha\mu} \delta_{\beta\nu} - H^{(\pm 1)}_{\nu\beta} \delta_{\alpha\mu} 
\\
(\alpha=\beta)
& \qquad
L^{(0)}_{\alpha\alpha;\mu\nu} =
\left( H^{(0)}_{\alpha\mu} \delta_{\alpha\nu} - H^{(0)}_{\nu\alpha} \delta_{\alpha\mu} \right)
+ i\hbar \left( \Gamma_{\mu\alpha} \delta_{\mu\nu} - \sum\limits_{q} \Gamma_{\alpha q}\delta_{\alpha\mu}\delta_{\alpha\nu} \right)
\\
(\alpha\neq\beta)
& \qquad
L^{(0)}_{\alpha\beta;\mu\nu} =
\left( H^{(0)}_{\alpha\mu} \delta_{\beta\nu} - H^{(0)}_{\nu\beta} \delta_{\alpha\mu} \right)
+ i\hbar \left( - \frac{1}{2} \sum\limits_{q} \left( \Gamma_{\alpha q} + \Gamma_{\beta q} \right) 
- \gamma_{\alpha\beta} \right) \delta_{\alpha\mu} \delta_{\beta\nu}
\end{array}
\end{equation}
\end{widetext}

The Hamiltonian and thus the Liouville supermatrix both oscillate at a frequency $\nu\equiv\Delta_2-\Delta_1$.  Therefore, the density supervector $\vec{\rho}(t)$ will have oscillatory components at frequencies that are harmonics of $\nu$.  We can use a Floquet expansion to express the supervector as a sum of slowly-varying supervector coefficients multiplied by oscillatory functions:
\begin{equation} \label{eq:FloquetRho}
\vec{\rho}(t) = \sum\limits_{m=-\infty}^{+\infty} \vec{\rho}^{(m)}(t) e^{i m \nu t}
\end{equation}
Substituting Eqns.~(\ref{eq:Lexpansion}) and (\ref{eq:FloquetRho}) into Eqn.~(\ref{eq:supervectorLiouville}) results in an infinite series of coupled linear differential equations for the supervector coefficients:
\begin{eqnarray}
\sum\limits_{m} \left( 
\frac{\partial\vec{\rho}^{(m)}}{\partial t} + i m \nu \vec{\rho}^{(m)}(t)
\right) e^{i m \nu t}
\nonumber \\
= - \frac{i}{\hbar}
\sum\limits_{np} L^{(n)} \vec{\rho}^{(p)} e^{i(n+p)t}
\end{eqnarray}

We invoke single-mode Floquet theory 
\cite{ho_floquet-liouville_1986}
to simplify this infinite series of equations.  To describe the rapidly oscillating factors, we define a Fourier state space $B_{(F)} = \{ \ket{\infty}, \ldots, \ket{1}, \ket{0}, \ket{-1}, \ldots, \ket{-\infty} \}$ where the state $\ket{m}$ represents oscillation at the $m^{\mathrm{th}}$ harmonic of the Hamiltonian oscillation frequency $\nu$:
\begin{equation}
\braket{t|m} = e^{i m \nu t}
\end{equation}
We also define operators on the Fourier space:
\begin{equation}
\begin{array}{ll}
F_z \ket{m} = m \ket{m}	&	F_z \equiv \sum\limits_{n=-\infty}^{+\infty} n \ket{n} \bra{n}  \\
F_m \ket{n} = \ket{n+m}	&	F_m \equiv \sum\limits_{n=-\infty}^{+\infty} \ket{n+m} \bra{n}
\end{array}
\end{equation}

The supermatrix $L(t)$ is an operator on the 16-dimensional Hilbert space defined by $B_{(H)} = \{ \ket{\alpha} \bra{\beta} \}$.  We define the Floquet space as the tensor product between the Hilbert space and Fourier space: $B^F = B_{(F)} \otimes B_{(H)}$.  Using Floquet space, we can express the finite-dimensional time-dependent Liouville supermatrix evolution equation  (\ref{eq:supervectorLiouville}) as an infinite-dimensional but time-\textit{independent} Floquet-Liouville supermatrix equation:
\begin{equation} \label{eq:FLSMevolution}
\frac{\partial}{\partial t} \vec{\rho}_F(t) = - \frac{i}{\hbar} L_F \vec{\rho}_F(t)
\end{equation}
where the Floquet-Liouville supermatrix is
\begin{equation} \label{eq:LFoperators}
L_F = \sum\limits_{n=-\infty}^{+\infty} \left( F_n \otimes L^{(n)} \right) 
	+ \hbar\nu \left( F_z \otimes I_{(H)} \right)
\end{equation}
and $I_{(H)}$ is the identity operator in Hilbert space.  Expressed as a matrix in Fourier space, $L_F$ is
\begin{widetext}
\begin{equation} \label{eq:LFmatrix}
L_F = 
\left[
\begin{matrix}
\ddots & \ddots & 0 & 0 & 0 & 0 & 0 \\
\ddots & L^{(0)}+2\nu I_{(H)} & L^{(1)} & 0 & 0 & 0 & 0 \\
0 & L^{(-1)} & L^{(0)}+\nu I_{(H)} & L^{(1)} & 0 & 0 & 0 \\
0 & 0 & L^{(-1)} & L^{(0)} &  L^{(1)} & 0 & 0 \\
0 & 0 & 0 & L^{(-1)} & L^{(0)}-\nu I_{(H)} &  L^{(1)} & 0 \\
0 & 0 & 0 & 0 & L^{(-1)} & L^{(0)}-2\nu I_{(H)} & \ddots \\
0 & 0 & 0 & 0 & 0 & \ddots & \ddots
\end{matrix}
\right]
\end{equation}
\end{widetext}
Each element of the above expression is a 16-by-16 Liouville supermatrix in the operator basis of Hilbert space.  The Floquet space density supervector is an infinite-dimensional vector:
\begin{equation}
\vec{\rho}_F (t) = \sum\limits_{m} \vec{\rho}^{(m)} (t) \ket{m}
= \sum\limits_{m\alpha\beta} \rho^{(m)}_{\alpha\beta} (t) \ket{m} \otimes \ket{\alpha} \bra{\beta}
\end{equation}
Expressed as a vector in Fourier space, $\vec{\rho}_F(t)$ is
\begin{equation}
\vec{\rho}_F(t) = 
\left[
\begin{matrix}
\vdots \\
\vec{\rho}^{(1)} (t) \\
\vec{\rho}^{(0)} (t) \\
\vec{\rho}^{(-1)} (t) \\
\vdots
\end{matrix}
\right]
\end{equation}
Because the Floquet-Liouville supermatrix in Eqns.~(\ref{eq:LFoperators}) and (\ref{eq:LFmatrix}) is time-independent, the solution to Eqn.~(\ref{eq:FLSMevolution}) is simple and well-known:
\begin{equation}
\vec{\rho}_F(t) = e^{-i L_F t / \hbar} \vec{\rho}_F (0)
\end{equation}
where $\vec{\rho}_F (0)$ is the initial Floquet supervector, which can be expressed in terms of the initial Liouville supervector, $\vec{\rho}(0)$, as
\begin{equation}
\vec{\rho}_F(0) = \left( \ket{0} \otimes \vec{\rho}(0) \right) = 
\left[
\begin{matrix}
\vdots \\
0 \\
\vec{\rho} (0) \\
0 \\
\vdots
\end{matrix}
\right]
\end{equation}

We can now determine the Floquet space density supervector at any arbitrary time simply by applying the evolution operator $U_F (t) \equiv e^{-i L_F t / \hbar}$ to the initial conditions $\vec{\rho}_F(0)$.  In the end, however, what is needed is the Hilbert space density supervector, which can be obtained from $\vec{\rho}_F(t)$ by applying the operator
\begin{equation}
\mathcal{O}_F(t) \equiv \sum\limits_{n} e^{i n \nu t} \bra{n} \otimes I_{(H)}
\end{equation}
The operator $\mathcal{O}_F (t)$ traces over just the Fourier space, leaving the Hilbert space density supervector.  It is an infinite row vector in Fourier space, whose elements are Hilbert space supermatrices:
\begin{widetext}
\begin{equation}
\mathcal{O}_F(t) = \left[
\begin{matrix}
\cdots & e^{i 2\nu t} I_{(H)} & e^{i \nu t} I_{(H)} & I_{(H)} & e^{-i \nu t} I_{(H)} & e^{-i 2\nu t} I_{(H)} & \cdots
\end{matrix}
\right]
\end{equation}
\end{widetext}
Finally, the Hilbert space density supervector at an arbitrary time with arbitrary initial conditions can be determined from the expression
\begin{equation} \label{eq:rhovec}
\vec{\rho}(t) = \mathcal{O}_F(t) U_F (t) \left( \ket{0} \otimes \vec{\rho} (0) \right)
\end{equation}

Though the supermatrices involved are in principle infinite-dimensional, in practice the Fourier space only needs a few dimensions, after which the space can be truncated without significantly altering the results of the computation.  The solution $\vec{\rho}(t)$ in Eqn.~(\ref{eq:rhovec}) can be calculated in a computationally efficient manner because it requires no numerical integration, only specification of the initial conditions and matrix algebra.

\subsection{\label{Simulations}Simulations}
We can now numerically calculate the density matrix evolution under experimental conditions.  A spin read-out operation will consist of resonant excitation in the pseudo-Faraday configuration where the QD is in one of the two electron-like eigenstates, $z+$  or $z-$.  Each electron-like eigenstate has one strong and one weakly-allowed optical transition.  To perform a projective measurement on the $z-$ eigenstate, for example, the resonant laser is tuned to the allowed $\sigma_{-}$ polarized transition (see Fig.~\ref{fig:Figure3}) and the fluorescence is detected.  Detection of a photon is interpreted as confirmation that the electron was in the $z-$ eigenstate.  The fidelity of the spin read-out depends on the specificity of the excitation and fluorescence.  If the resonant laser is tuned to the $\sigma_{-}$ polarized transition, but the electron is in the $z+$  spin state, photon emission may still be detected, giving an erroneous signal.

To predict the fidelity of the spin read-out operation, we simulate it using typical QD parameters and a linearly-polarized resonant laser tuned to the $\sigma_{-}$ cycling transition. The experimentally controllable parameters used are: $B_x= 0.1$ T, $\Omega_{1+}/2\pi = 200$ GHz, $\Delta_1/2\pi = 2$ THz, $\Omega_{2+}/2\pi = \Omega_{2-}/2\pi = 0.5$ GHz.
Note that the AC Stark mixing parameter is kept small $(\Omega_{1+}/2\Delta_1 = 0.05)$, but the AC Stark shift is still large: $\Omega_{1+}^2/2\Delta_1 = 2\pi(10 \mathrm{GHz})$.  In order to have the excitation laser in resonance with the $\sigma_{-}$ cycling transition, the detuning $\Delta_2$ is chosen to match the difference between the eigenvalues of the time-independent pseudo-Faraday Hamiltonian $H^{(0)}$ that correspond to the $z-$ electron-like and trion-like states.  The QD parameters would not be controllable in practice, except by the choice of QD, but typical values are used here.  The electron and hole g-factors \cite{kroner_resonant_2008} are $g_{e,x}= 0.24$, $g_{h,x}= 0.47$.  The population relaxation rates $\Gamma_{\alpha\beta}$ used in the simulation are all zero except for: the spontaneous emission rates 
\cite{flagg_dynamics_2012}
$\Gamma_{31} = \Gamma_{42} = 1.54$ GHz;
the weakly allowed emission rates
\cite{delteil_observation_2014}
$\Gamma_{41} = \Gamma_{32} = 3.42$ MHz;
and the electron spin decay rates
\cite{kroutvar_optically_2004}
$\Gamma_{21} = \Gamma_{12} = 50$ Hz.
The dephasing rates $\gamma_{\alpha\beta}$ used are: trion dephasing
\cite{flagg_interference_2010}
$\gamma_{31}=\gamma_{13}=\gamma_{42}=\gamma_{24}=1.72$ GHz;
electron spin dephasing
\cite{xu_optically_2009}
$\gamma_{12}=\gamma_{21}= 12.6$ MHz.

We numerically calculate the density matrix evolution for two initial conditions: the $z-$ and $z+$  electron-like states.  The photon emission rate is proportional to the population in the trion states and is given by:
\begin{equation}
R(t) = \Gamma_{31} \rho_{33}(t) + \Gamma_{42} \rho_{44}(t)
\end{equation}
The average number of detected photons is the overall detection efficiency $\epsilon$ multiplied by the time integral of $R(t)$ over the duration $T$ of the detection window:
\begin{equation}
D(T) = \epsilon \int_{0}^{T} R(t) dt
\end{equation}

\begin{figure}[t]
\includegraphics{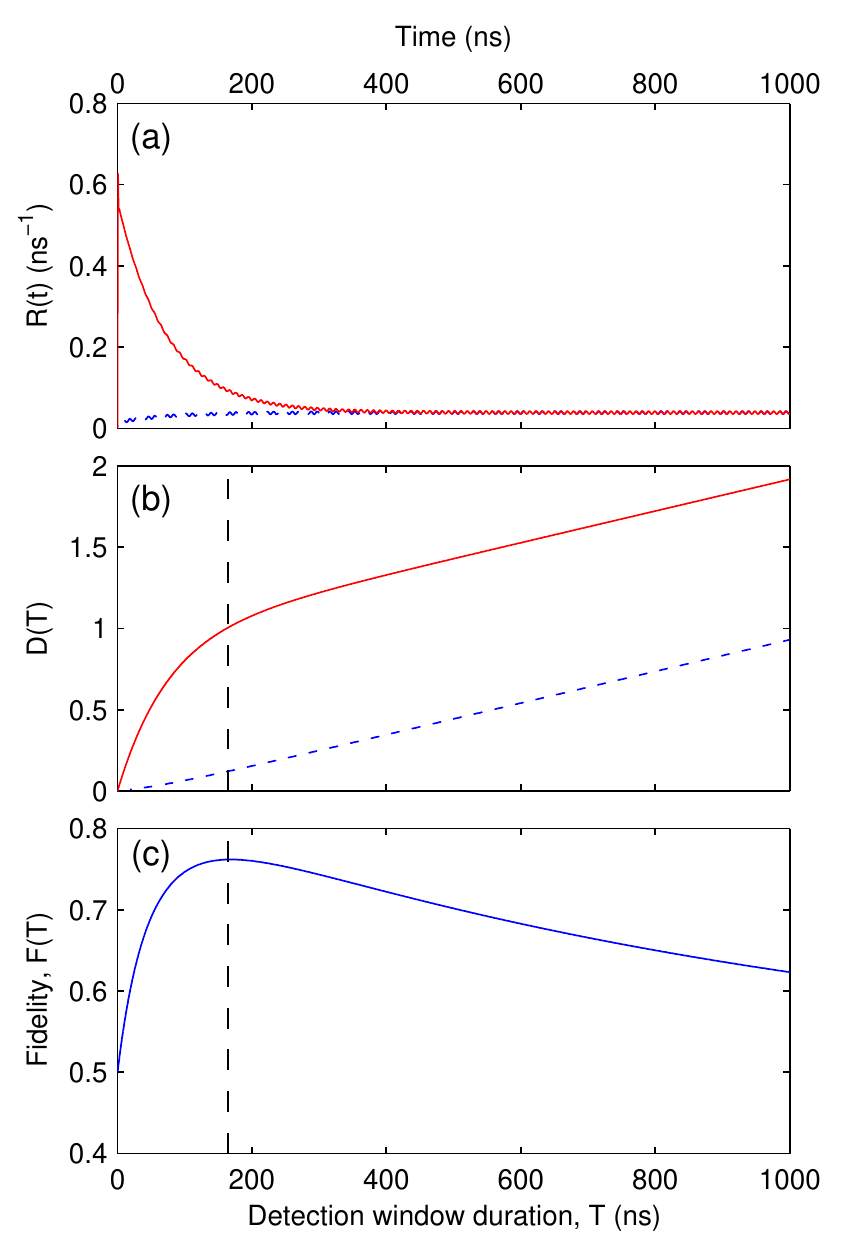}
\caption{\label{fig:Figure5} (Color online) Spin read-out operation under $\sigma_{-}$ resonant excitation.  In (a) and (b) the two initial conditions are the $z-$ electron state (solid red curve) and the $z+$ electron state (dashed blue curve). (a) Photon emission rate after $\sigma_{-}$ resonant excitation begins. (b) Average number of photons detected as a function of detection window duration. (c) Fidelity of the spin measurement as a function of detection window duration.  The vertical dashed line indicates the optimum detection window.}
\end{figure}

Figure \ref{fig:Figure5}(a) shows the photon emission rate $R(t)$ as a function of time after resonant excitation begins.  When the initial state is the $z-$ electron state (solid red curve) the emission starts strong and decays as the population is pumped into the $z+$  state, which is not being resonantly excited. When the initial state is the $z+$  electron state (dotted blue curve) the emission rate starts very low but rises slightly as a small amount of population is pumped into the $z-$ electron state.  Figure \ref{fig:Figure5}(b) shows the average number of photons detected $D(T)$ as a function of the detection window duration for the two initial conditions. When the initial state is the $z+$  electron state (dotted blue curve), photon emission is relatively unlikely because the excitation laser is not resonant with the allowed $\sigma_{+}$ transition; emission is still possible, however, due to weak remaining magnetic spin mixing. When the initial state is the $z-$ electron state (solid red curve), photon emission is relatively likely.  The fidelity of the spin measurement is 
$F = (1-p_{z+} + p_{z-}) / 2$
where $p_{z\pm}$ is the probability of detecting at least one photon when the initial state is $z\pm$.  The fidelity is plotted in Fig.~\ref{fig:Figure5}(c) as a function of detection window duration.  The maximum fidelity occurs for a detection window of 165 ns, indicated by a vertical dashed line, beyond which the value of $p_{z-}$ saturates but $p_{z+}$ keeps increasing.  For the optimum detection window duration the calculated fidelity is 76.2\%, which is slightly lower than the 82.3\% measured in the true Faraday configuration in 
Ref.~[\onlinecite{delteil_observation_2014}].
The fidelity would be improved with larger AC Stark shift, smaller Voigt magnetic field, smaller electron spin dephasing and decay rates, or smaller trion dephasing rate.  The values chosen here are either typical or feasible for real experiments.

When the average number of photons detected is greater than unity, we can say that a single-shot measurement of the electron spin-state is possible 
\cite{delteil_observation_2014}.
For these calculations, we have chosen an overall detection efficiency $\epsilon=2.5\%$ that results in an average number of detected photons greater than 1 for the optimum detection window duration: for $T=165$ ns, $D(T)=1.01$.  Such an efficiency is relatively high, but should be achievable in recently developed ``photonic trumpet'' waveguides 
\cite{munsch_dielectric_2013},
which have a demonstrated first-lens collection efficiency of 75\%.  Because application of a non-resonant laser is necessary for the AC Stark effect, a sample design such as the trumpet that achieves high collection efficiency through non-resonant effects is necessary, which eliminates most microcavities from consideration.

\section{\label{Conclusion}Conclusion}
We have described a scheme to accomplish a single-shot read-out of the spin-state of an electron trapped in a quantum dot while maintaining the capability to perform arbitrary coherent manipulation of the spin-state.  A pseudo-Faraday configuration is produced by application of a Voigt geometry magnetic field and a far red-detuned, circularly-polarized laser that causes a spin-dependent AC Stark shift.  The spin read-out is accomplished in the pseudo-Faraday configuration {\it via} spin-selective fluorescence from spin-preserving cycling transitions.  For typical quantum dot parameters and feasible detection efficiency, the spin measurement can be accomplished faster than the state is disturbed by the back-action, resulting in a single-shot read-out.  The fidelity of the read-out is limited by the remaining spin-state mixing in the pseudo-Faraday configuration that is caused by the Voigt geometry magnetic field.  Because the laser that produces the AC Stark shift can be switched on and off rapidly compared to the spin lifetime, this scheme offers the possibility to perform coherent spin manipulation in the Voigt configuration and then single-shot spin read-out in the pseudo-Faraday configuration.  These capabilities comprise all three necessary single-qubit operations and will allow the investigation of more complex control and manipulation sequences.

\begin{acknowledgments}
The authors wish to acknowledge helpful discussions with J. M. Taylor and A. Tudorascu.
\end{acknowledgments}

\appendix*
\section{Branching ratio}
The branching ratio for the pseudo-Faraday configuration can be derived by considering the four-level charged QD system interacting with the quantized multi-mode electromagnetic field.  Extending the derivation in reference [\onlinecite{scully_quantum_1997}] to four levels, the state of the system as a function of time is
\begin{align} \label{eq:Psi_definition}
\ket{\Psi(t)} =& C_4(t) \ket{\psi_4, \mathrm{vac}} + C_3(t) \ket{\psi_3, \mathrm{vac}} \nonumber \\
	+ &\sum_{\mathbf{k}} \left\{ C_{2 \mathbf{k}}(t) \ket{\psi_2, \mathbf{k}} + C_{1 \mathbf{k}}(t) \ket{\psi_1,\mathbf{k}} \right\}
\end{align}
and the interaction Hamiltonian in the rotating wave approximation is
\begin{equation} \label{eq:interaction_Hamiltonian}
\mathcal{V} = \hbar \sum_{\mathbf{k}} \sum_{\overset{i=1,2}{j=3,4}} \left\lbrace g_{ij,\mathbf{k}} \sigma_{ij} a^{\dagger}_{\mathbf{k}} e^{-i\left( \omega_{ji} - \omega_{k} \right) t} + H.c. \right\rbrace
\end{equation}
where the $\ket{\psi_i}$ are the eigenstates of the system, $\sigma_{ij} = \ket{\psi_i}\bra{\psi_j}$ is the QD lowering operator, $a_{\mathbf{k}}$ is the photon annihilation operator for the electromagnetic field mode with wavevector $\mathbf{k}$ and frequency $\omega_{k}$, $\omega_{ji}$ is the transition frequency between states $\ket{\psi_j}$ and $\ket{\psi_i}$, $g_{ij,\mathbf{k}}$ is a coupling constant, and $H.c.$ means the Hermitian conjugate.  The coupling constant is given by
\begin{equation} \label{eq:coupling_constant}
g_{ij,\mathbf{k}} \equiv - \vec{\mathcal{P}}_{ij} \cdot \hat{\epsilon}_{\mathbf{k}} \mathcal{E}_{\mathbf{k}} / \hbar
\end{equation}
where $\vec{\mathcal{P}}_{ij} \equiv e \bra{\psi_i} \mathbf{r} \ket{\psi_j}$ is the electric dipole matrix element in the eigenbasis and in general is a complex vector, $\hat{\epsilon}_{\mathbf{k}}$ is the polarization vector of the $\mathbf{k}$-mode, and 
\begin{equation} \label{eq:fieldamplitude}
\mathcal{E}_{\mathbf{k}} = \left( \frac{\hbar \omega_k}{2 \epsilon_0 V}  \right) ^{1/2}
\end{equation}
where $\epsilon_0$ is the permittivity of free space, and $V$ is the quantization volume.

To determine the time evolution of the amplitude coefficients of $\ket{\Psi(t)}$ we substitute Eqn.~(\ref{eq:Psi_definition}) into the interaction picture Schr\"odinger equation and equate the coefficients of similar kets.  The result is a series of coupled linear differential equations:
%
%\begin{eqnarray}
%\dot{C}_4 &=& -i \sum_{\mathbf{k}} e^{-i \omega_{k} t} \left( g^{*}_{14,\mathbf{k}} e^{i \omega_{41} t} C_{1\mathbf{k}} + g^{*}_{24,\mathbf{k}} e^{i \omega_{42} t} C_{2\mathbf{k}} \right) \label{eq:C4_EoM} \\
%\dot{C}_3 &=& -i \sum_{\mathbf{k}} e^{-i \omega_{k} t} \left( g^{*}_{13,\mathbf{k}} e^{i \omega_{31} t} C_{1\mathbf{k}} + g^{*}_{23,\mathbf{k}} e^{i \omega_{32} t} C_{2\mathbf{k}} \right) \label{eq:C3_EoM} \\
%\dot{C}_{2\mathbf{k}} &=& -i \sum_{\mathbf{k}} e^{i \omega_{k} t} \left( g_{24,\mathbf{k}} e^{-i \omega_{42} t} C_4 + g_{23,\mathbf{k}} e^{-i \omega_{32} t} C_3 \right) \label{eq:C2k_EoM} \\
%\dot{C}_{1\mathbf{k}} &=& -i \sum_{\mathbf{k}} e^{i \omega_{k} t} \left( g_{14,\mathbf{k}} e^{-i \omega_{41} t} C_4 + g_{13,\mathbf{k}} e^{-i \omega_{31} t} C_3 \right) \label{eq:C1k_EoM}
%\end{eqnarray}
%
%
\begin{widetext}
\begin{eqnarray}
\dot{C}_4 &=& -i \sum_{\mathbf{k}} \left( g^{*}_{14,\mathbf{k}} e^{i \left( \omega_{41} - \omega_k \right) t} C_{1\mathbf{k}} + g^{*}_{24,\mathbf{k}} e^{i \left( \omega_{42} - \omega_k \right) t} C_{2\mathbf{k}} \right) \label{eq:C4_EoM} \\
\dot{C}_3 &=& -i \sum_{\mathbf{k}} \left( g^{*}_{13,\mathbf{k}} e^{i \left( \omega_{31} - \omega_k \right) t} C_{1\mathbf{k}} + g^{*}_{23,\mathbf{k}} e^{i \left( \omega_{32} - \omega_k \right) t} C_{2\mathbf{k}} \right) \label{eq:C3_EoM} \\
\dot{C}_{2\mathbf{k}} &=& -i \sum_{\mathbf{k}} \left( g_{24,\mathbf{k}} e^{-i \left( \omega_{42} - \omega_k \right) t} C_4 + g_{23,\mathbf{k}} e^{-i \left( \omega_{32} - \omega_k \right) t} C_3 \right) \label{eq:C2k_EoM} \\
\dot{C}_{1\mathbf{k}} &=& -i \sum_{\mathbf{k}} \left( g_{14,\mathbf{k}} e^{-i \left( \omega_{41} - \omega_k \right) t} C_4 + g_{13,\mathbf{k}} e^{-i \left( \omega_{31} - \omega_k \right) t} C_3 \right) \label{eq:C1k_EoM}
\end{eqnarray}
\end{widetext}
These equations can be solved following the usual Weisskopf-Wigner theory \cite{scully_quantum_1997}.  First, we directly integrate Eqns.~(\ref{eq:C2k_EoM}) and (\ref{eq:C1k_EoM}) and substitute them into Eqns.~(\ref{eq:C4_EoM}) and (\ref{eq:C3_EoM}).  Then we assume that the $\mathbf{k}$-modes are closely spaced, which changes the sum over $\mathbf{k}$ into an integral over $\mathbf{k}$-space.  The result is two coupled linear differential equations for $C_4$ and $C_3$, one of which is
\begin{widetext}
\begin{align} \label{eq:C4_ODE_w_dotproduct}
\dot{C}_4 =& - \frac{2V}{(2\pi)^3 \hbar^2}
\int d^3k \mathcal{E}_{\mathbf{k}}^2 \int_0^t dt'  \nonumber \\
\left\lbrace \vphantom{\int_0} \right.
	&\left| \vec{\mathcal{P}}_{14} \cdot \hat{\epsilon}_{\mathbf{k}} \right| ^2
	e^{i (\omega_{41} - \omega_k) (t-t') } C_4(t')
	%\nonumber \\
	+\left( \vec{\mathcal{P}}_{14}^* \cdot \hat{\epsilon}_{\mathbf{k}} \right) \left( \vec{\mathcal{P}}_{13} \cdot \hat{\epsilon}_{\mathbf{k}} \right)
	e^{i (\omega_{41} - \omega_k) t } e^{-i (\omega_{31} - \omega_k) t' } C_3(t')
	\nonumber \\
	+&\left| \vec{\mathcal{P}}_{24} \cdot \hat{\epsilon}_{\mathbf{k}} \right| ^2
	e^{i (\omega_{42} - \omega_k) (t-t') } C_4(t')
	%\nonumber \\
	+\left( \vec{\mathcal{P}}_{24}^* \cdot \hat{\epsilon}_{\mathbf{k}} \right) \left( \vec{\mathcal{P}}_{23} \cdot \hat{\epsilon}_{\mathbf{k}} \right)
	e^{i (\omega_{42} - \omega_k) t } e^{-i (\omega_{32} - \omega_k) t' } C_3(t')
\left. \vphantom{\int_0}\right\rbrace
\end{align}
\end{widetext}

Each term of the integrand in Eqn.~(\ref{eq:C4_ODE_w_dotproduct}) has a factor of the form
\begin{equation}
\left( \mathbf{u} \cdot \hat{\epsilon}_{\mathbf{k}} \right)  \left( \mathbf{v} \cdot \hat{\epsilon}_{\mathbf{k}} \right)
\end{equation}
where e.g., $\mathbf{u} = \vec{\mathcal{P}}_{14}^*$ and $\mathbf{v} = \vec{\mathcal{P}}_{13}$.  Since $\hat{\epsilon}_{\mathbf{k}}$ is a unit vector these factors only depend on the angular part of the integral over $\mathbf{k}$-space.  We define a general integral as follows:
\begin{equation} \label{eq:dotproduct_integral}
A \equiv \int d^2 \Omega_{\mathbf{k}} \left( \mathbf{u} \cdot \hat{\epsilon}_{\mathbf{k}} \right)  \left( \mathbf{v} \cdot \hat{\epsilon}_{\mathbf{k}} \right)
\end{equation}
which can be directly integrated to give
\begin{equation} \label{eq:general_dotproduct_factor}
A = \frac{4\pi}{3} \mathbf{u} \cdot \mathbf{v}
\end{equation}
Thus, the angular parts of the $\mathbf{k}$-space integral in Eqn.~(\ref{eq:C4_ODE_w_dotproduct}) can be replaced using Eqn.~(\ref{eq:general_dotproduct_factor}) with appropriate substitutions.  We can continue the Weisskopf-Wigner theory with the approximation that the coefficients $C_4(t)$ and $C_3(t)$ evolve much slower than the oscillation frequency $\omega$ (see reference [\onlinecite{scully_quantum_1997}] for details).  This allows us to perform the integrals over $k$ and $t'$ in Eqn.~(\ref{eq:C4_ODE_w_dotproduct}) to obtain
\begin{align} \label{eq:C4_ODE_tprime_integral_done}
\dot{C}_4 =& - \frac{1}{2} \left( \Gamma_{41} + \Gamma_{42} \right) C_4(t)
	- \frac{1}{2} \left( \beta_{31} + \beta_{32} \right) e^{i \omega_{43} t} C_3(t)
\end{align}
where the spontaneous decay rates $\Gamma_{41}$ and $\Gamma_{42}$ are
\begin{equation}
\Gamma_{ji} \equiv \frac{1}{4\pi\epsilon_0} \frac{4 \omega_{ji}^3}{3\hbar c^3} \left| \vec{\mathcal{P}}_{ij} \right|^2
\end{equation}
and the transition rates $\beta_{31}$ and $\beta_{32}$ are
\begin{equation}
\beta_{ji} \equiv \frac{1}{4\pi\epsilon_0} \frac{4 \omega_{ji}^3}{3\hbar c^3} \left( \vec{\mathcal{P}}_{il}^* \cdot \vec{\mathcal{P}}_{ij} \right)
\end{equation}
where $l=4$ when $j=3$ and vice versa.  An expression similar to Eqn.~(\ref{eq:C4_ODE_tprime_integral_done}) can be obtained for $\dot{C}_3$ by switching all the indices 3 and 4.

The transition rates $\beta_{ji}$ are zero when the dipole moments for the two transitions to one electron state are orthogonal.  This is the case for a charged self-assembled QD in the Faraday, pseudo-Faraday, or Voigt configurations, which can be confirmed using Eqn.~(\ref{eq:dipole_moment}) and the eigenvectors of $H_0$ from Eqn.~(\ref{eq:H0matrix}).  Thus Eqn.~(\ref{eq:C4_ODE_tprime_integral_done}) contains only the spontaneous decay rates.  When the dipole moments are non-orthogonal, there is the possibility of spontaneously generated coherence \cite{economou_unified_2005,dutt_stimulated_2005} and/or quantum interference causing population transfer from one trion state to the other \cite{zhu_spontaneous_1995,plastina_spontaneous_1999}.

The branching ratio is the ratio of the spontaneous decay rates $\Gamma_{14}$ and $\Gamma_{24}$.  The difference between the transition frequencies $\omega_{41}$ and $\omega_{42}$ is much smaller than their magnitude, so we can approximate the branching ratio $r_B$ as
\begin{equation} \label{eq:branching_ratio_w_dipole_moments}
r_B = \frac{ \left| \vec{\mathcal{P}}_{14} \right|^2 }{ \left| \vec{\mathcal{P}}_{24} \right|^2 }
\end{equation}

From the definition of $\vec{\mathcal{P}}_{ij}$, above, we can use the completeness of the Faraday basis $\left\{ \ket{\alpha} \right\}$ from Eqns.~(\ref{eq:FaradayBasis}) to obtain
\begin{equation}  \label{eq:dipole_as_matrix_product}
\vec{\mathcal{P}}_{ij} = 
	\sum_{\alpha\beta} \braket{\psi_i | \alpha} \braket{\beta | \psi_j} \mathbf{q}_{\alpha\beta}
\end{equation}
where $\mathbf{q}_{\alpha\beta} \equiv e \bra{\alpha} \mathbf{r} \ket{\beta}$ is the electric dipole vector matrix element in the Faraday basis.  Note that $\mathbf{q}_{\alpha\beta} = \mathbf{q}_{\beta\alpha}^*$ and some of the $\mathbf{q}_{\alpha\beta}$ are zero.  For example, $\mathbf{q}_{\alpha\alpha}=0$ by parity for all values of $\alpha$.  Also, $\mathbf{q}_{43} = \mathbf{q}_{21} = 0$ at optical frequencies.  We know that the polarization selection rules of the Faraday configuration are such that the allowed transitions are $\ket{1} \leftrightarrow \ket{3}$ and $\ket{2} \leftrightarrow \ket{4}$.  Thus, we can assert that only 4 of the remaining $\mathbf{q}_{\alpha\beta}$ are non-zero: $\mathbf{q}_{13}$, $\mathbf{q}_{31}$, $\mathbf{q}_{24}$, and $\mathbf{q}_{42}$.  We know that in the Faraday configuration the allowed transitions are circularly polarized when the photon is emitted in the $z$-direction.  The dipole moments for such transitions are
\begin{equation} \label{eq:defining_qmn}
\begin{array}{l}
\mathbf{q}_{13} = \frac{1}{\sqrt{2}} \left( \hat{x} + i \hat{y} \right) \\
\mathbf{q}_{24} = \frac{1}{\sqrt{2}} \left( \hat{x} - i \hat{y} \right)
\end{array}
\end{equation}

Combining Eqns.~(\ref{eq:dipole_as_matrix_product}), (\ref{eq:defining_qmn}), and (\ref{eq:QD_lowering_operators}) we can define the electric dipole operator in the Faraday basis as
\begin{equation} \label{eq:dipole_moment_operator}
\mathbf{d} = \mathbf{q}_{13} \sigma_+  +  \mathbf{q}_{24} \sigma_-  + H.c.
\end{equation}
and the transition dipole moments between the eigenstates are
\begin{equation} \label{eq:dipole_moment}
\vec{\mathcal{P}}_{ij} = \bra{\psi_i} \mathbf{d} \ket{\psi_j}
\end{equation}
Thus, the branching ratio in Eqn.~(\ref{eq:branching_ratio}) can be derived by combining Eqns.~(\ref{eq:branching_ratio_w_dipole_moments}) and (\ref{eq:dipole_moment}).

%\newcommand{\Library}{../Figures-Bib/Library}
\bibliography{Library}

\end{document}